\documentclass[useAMS,usenatbib]{mn2e}
\usepackage{epsfig}
\usepackage{subfigure}
\usepackage{bm}

\def\lsim{\mathrel{\rlap{\lower3.5pt\hbox{\hskip0.5pt$\sim$}}
    \raise0.5pt\hbox{$<$}}}
\def\gsim{~\rlap{$>$}{\lower 1.0ex\hbox{$\sim$}}}

\title[Slope evolution of GRB correlations and cosmology]{Slope evolution of GRB correlations and cosmology}

\author[M.G. Dainotti et al.]{M.G. Dainotti$^{1,2}$\thanks{Corresponding author\,: {\tt mdainott@stanford.edu;\newline
mariagiovannadainotti@yahoo.it}}, V.F. Cardone$^{3}$, E. Piedipalumbo $^{4,5}$, S. Capozziello$^{4,5}$ \\
$^1$Astronomy Department,Stanford University, Via Pueblo Mall 382, Stanford, CA, USA\\
$^2$Obserwatorium Astronomiczne, Uniwersytet Jagiello\'{n}ski, ul. Orla 171, 30-244 Krak\'{o}w, Poland \\
$^3$I.N.A.F.\,-\,Osservatorio Astronomico di Roma, via Frascati 33, 00040 - Monte Porzio Catone (Roma), Italy \\
$^4$Dipartimento di Fisica, Universit\`{a} di Napoli Federico II, Compl. Univ. Monte S. Angelo, 80126 Naples, Italy \\
$^5$I.N.F.N., Sez. di Napoli, Compl. Univ. Monte S. Angelo, Edificio G, via Cinthia, 80126 - Napoli, Italy \\}

\date{Accepted xxx, Received yyy, in original form zzz}

\begin{document}

\maketitle

\begin{abstract}

 Gamma -ray bursts (GRBs) observed up to redshifts $z>9.4$ can be used as possible probes to test cosmological models. Here we show how changes of the slope of the {\it luminosity $L^*_X$ -break time $T^*_a$} correlation in GRB afterglows, hereafter the LT correlation, affect the determination of the cosmological parameters. 
With a simulated data set of 101 GRBs with a central value of the correlation slope that differs on the intrinsic one by a $5\sigma$ factor, we find an overstimated value of the matter density parameter, $\Omega_M$, compared to the value obtained with SNe Ia, while the Hubble constant, $H_0$, best fit value is still compatible in 1$\sigma$ compared to other probes.
We show that this compatibility of $H_0$ is due to the large intrinsic scatter associated with the simulated sample. Instead, if we consider a subsample of high luminous GRBs ($HighL$), we find that both the evaluation of $H_0$ and $\Omega_M$ are not more compatible in 1$\sigma$ and $\Omega_M$ is underestimated by the $13\%$.
However, the $HighL$ sample choice reduces dramatically the intrinsic scatter of the correlation, thus possibly identifying this sample as the standard canonical `GRBs' confirming previous results presented in Dainotti et al. (2010,2011). Here, we consider the LT correlation as an example, but this reasoning can be extended also for all other GRB correlations. In literature so far GRB correlations are not corrected for redshift evolution and selection biases, therefore we are not aware of their intrinsic slopes and consequently how far the use of the observed correlations can influence the derived `best' cosmological settings.
Therefore, we conclude that any approach that involves cosmology should take into consideration only intrinsic correlations not the observed ones.

\end{abstract}


\section{Introduction}

The high fluence values (from $10^{-7}$ to $10^{-4} \ {\rm  erg/cm^2}$) and the huge isotropic energy, $E_{iso}$, emitted ($\simeq 10^{48} - 10^{54} {\rm erg}$) in
the prompt emission phase make GRBs the most violent and energetic astrophysical phenomena. These same features allow to detect them up to very high redshift thus offering the intriguing possibility to use them as standard candles to trace the Hubble diagram deep into the matter dominated era. To this end, one has to rely on scaling relations between an observable redshift independent quantity and a distance dependent one so that the measurement of the former allows the determination of the distance. Many empirical motivated correlations are presently available to carry on this program \citep{Amati08,FRR00,N00,G04,liza05} thus fueling the hope to turn GRBs into standardizeable distance indicators as Type Ia Supernovae (hereafter, SNeIa). However, the above correlations, have as one variable $E_{iso}$ and because of that they suffer of a double truncation due to detection selection threshold \citep{Lloyd1999}.
Notwithstanding this problem, these correlations have been used to constrain cosmological models. Combining the estimates from different correlations, Schaefer (2007) first derived the GRBs Hubble diagram (hereafter, HD) for 69 objects, while Cardone et al. (2009, 2010) used a different calibration method and add the Luminosity-time correlation \citep{DCC08,Dainotti2010,D11} to update the GRBs HD. A more recent compilation of GRBs with measured values of different correlation related quantities has been presented in \cite{XS10} and used in \cite{Marcy} to investigate the impact of systematics on the GRBs HD.

Notwithstanding these remarkable first attempts, whether GRBs can indeed be considered standardizeable distance indicators is a still pending question undergoing a fierce debate. To this regard, it is worth remembering that a well behaved distance indicator should be not only visible to high $z$ and possess scaling relations with as small as possible intrinsic scatter, but its physics should be well understood from a theoretical point of view \citep{Petrosian2009}. Moreover, so far GRBs scaling relations are used as cosmological tool, adopting the observables of the raw correlations not corrected for redshift evolution and selection biases.

A valid tool in classifying GRBs is provided by the analysis of their light curves. A crucial breakthrough in this field has been represented by the launch of the {\it Swift} satellite in 2004, which allows a rapid follow\,-\,up of the afterglows in different wavelengths giving better coverage of the GRB light curve than the previous missions. Such data revealed the existence of a more complex phenomenology with different slopes and break times thus stressing the inadequacy of a single power\,-\,law function. A significant step forward has been made by the analysis of the X\,-\,ray afterglow curves of the full sample of {\it Swift} GRBs showing that they may be fitted by a single analytical expression \citep{W07} which we referred to in the following as the W07 model.

\cite{DCC08} first found that the break time $T_a^* = T_a/(1 + z)$ and the luminosity at the break time $L_X^*$ (where $z$ is the GRB redshift and the asteriks refer to the rest frame quantities) are not independent, but rather follow the log\,-\,linear relation, $\log L^*_X =  a \log T^*_a + b$, with $a$ and $b$ fixed by the D'Agostini fitting method \citep{Dago05}. Such correlation has been first confirmed \citep{Ghisellini2008,Yamazaki09} and then updated with 77 GRBs \citep{Dainotti2010}, while the possible impact of systematics on the small error parameters sample have been investigated in \cite{Dainotti2011b}. Recently, Dainotti et al. 2013 have demonstrated the intrinsic nature of the LT correlation and provided an evaluation of the time and luminosity evolution.
Moving along this route, we present here an estimation of how much the evaluation of the cosmological parameters is biased if we use an LT correlation slope that depart from the intrinsic correlation slope of 5 $\sigma$. Moreover, we address the calibration problem relying on an improved Bayesian analysis which explicitly takes into account the uncertainties on the background cosmological parameters.

The plan of the paper is the following: in Sect.\,2, we present the simulated dataset we use and the intrinsic LT correlation; sect.\,3 is devoted to the description of the statistical method used to jointly determine both the calibration parameters and the background cosmological model. Results are summarized and discussed in Sect.\,4, while conclusions are presented in Sect.\,5.

\section{The simulated  LT correlation and motivation for the chosen subsample}

Since some of the LT correlation could result from the redshift dependences of the observables, Dainotti et al. (2013) use the Efron-Petrosian (1992), hereafter EP method to reveal the intrinsic nature of this correlation. 
To make the reader more acquainted on the reason why here we depart from the intrinsic slope we discuss briefly the results obtained in this paper and its consequences.
For the updated sample of 101 GRBs, namely the whole sample without the short bursts with extended emission \citep{nb2010}, called the Intermediate class (IC), Dainotti et al. 2013 found the power law slope $a=-1.27 _{-0.26}^{+0.18}$, while for the whole sample $a=-1.32 _{-0.17}^{+0.18}$ with the D'Agostini method.
Therefore, we note a steepening of the slope enlarging the sample from the past years \citep{DCC08,Dainotti2010}.
  The Spearman correlation coefficient for the larger sample ($\rho=-0.74$) is higher than $\rho=-0.68$ obtained for a subsample of 66 long duration GRBs analyzed in Dainotti et al. 2010. The probability of the correlation (of the 101 long GRBs) occurring by chance within an uncorrelated sample is $P \approx 10^{-18}$ \citep{Bevington}.

To determine the intrinsic slope of the correlation it is necessary to evaluate whether
the variables $L_X$ and $T^*_a$, are correlated with redshift or are
statistically independent of it. For example, the correlation between $L_X$ and the redshift, $z$, is what we
call luminosity evolution, and independence of these variables would imply absence of such evolution. The EP method prescribed how to remove the correlation by defining
new and independent variables.

\begin{figure}
\includegraphics[width=7 cm]{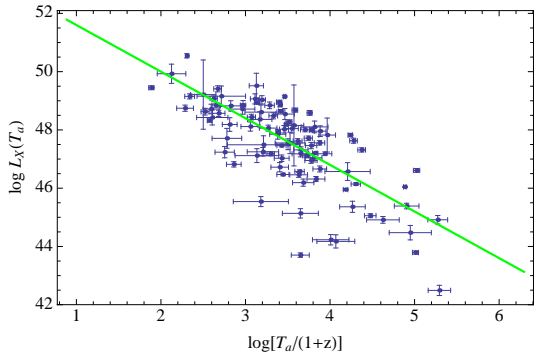}
\caption{Best fit curves for the  $ L_X^{*}$ -- $ T_a^{*}$ correlation relation superimposed on
the data in our  full bayesian approach, as it results for the FlatCPL model and GRB\,+\,SNeIa\,+\,$H(z)$\,+\,$H_0$ datasets.  }
\label{lxtaplot}
\end{figure}

Dainotti et al. 2013 determined the correlation functions, $g(z)$ and $f(z)$ when determining the evolution of $L_X$ and $T^{*}_a$ so that de-evolved variables $L'_{X} \equiv L_X/g(z)$ and $T'_a \equiv T^*_a/f(z)$ are not correlated with z.
The evolutionary function are parametrized by simple correlation functions

\begin{equation}
g(z)=(1+z)^{k_{Lx}}, f(z)=(1+z)^{k_{T^{*}a}}
\label{lxev}
\end{equation}

so that $L'_{X}=L_X/g(z)$ refer to the local ($z=0$) luminosities. Dainotti et al. 2013 found that there is no discernable luminosity evolution, $k_{L_x}=-0.05_{-0.55}^{+0.35}$, but there is an evolution in $T^{*}_a$, $k_{T^{*}a}=-0.85_{-0.30}^{+0.30}$ especially at high redshift.
Moreover, applying again the EP method to the de-evolved observables, they found that the correlation between $L^{'}_X$ and $T^{'}_a$ is $a=-1.07_{-0.14}^{+0.09}$.
Therefore, we note a steepening in the slope parameters in the real data, when we consider a previous analysis \citep{Dainotti2010} and in Dainotti et al. 2013 we attribute this steepening to selection effects and redshift evolution of the observables.
Therefore, the aim of the paper is to show how much a departure from 5 $\sigma$ above and below the central value of the intrinsic slope can affect the cosmological results. There is a wide discussion about the reliability of the GRB as possible cosmological probes due to the wide scatter of GRB correlations and due to the fact that all the GRB correlations may be affected by redshift evolution of their parameters and selection effects. Therefore, it is important to put a limit on how far we can depart from an intrinsic distribution of observables before changing dramatically the estimate of cosmological models.
Another question we address is how much difference exists between results obtained from a subsample of high luminous GRBs and the ones of the overall sample.
We discuss here a simulated data set of 101 GRBs with an imposed correlation slope, $a=-1.52$ assuming the fiducial $\Lambda$CDM flat cosmological model with $\Omega_M = 0.291$ and $H_0 = 71  {\rm Km s}^{-1} {\rm Mpc}^{-1}$, see Fig. \ref{lxtaplot}. We built this set associating to each real data of the distribution a simulated $L_X$ with the slope $-1.52$ obtained performing a Monte Carlo simulation with the real distributions. More precisely, the simulated luminosities are determined by applying an LT correlation with $\log L_X \approx -1.52 \log T^{*}_a$, where $-1.52$ is the imposed $a$ slope of LT correlation and we chose as intrinsic scatter $\sigma_{int}=0.93$, much wider than the scatter obtained from the real observed data sample, $\sigma_{int}=0.66$. The aim of increasing the scatter is to see how much with a wider dispersion the best fit parameters of the cosmology change and how much this scatter can be reduced if we consider the high luminous subsample.
From this total sample of GRBs we selected a subsample, called {\em HighL} samples, with the condition that $L_X \geq 48.7$ and the other sample will be called {\em Full}.

The reason for this choice is that we have demonstrated in Dainotti et al. 2013 that the luminosity function corrected by the redshift evolution and selection effects is equal to the observed luminosity function for luminosities $L_X \geq 48$. Here we could have chosen a sample with this exact feature, but in such a case we would have had a smaller (than 5 $\sigma$) difference from the intrinsic slope. We keep a symmetry of a 5 $\sigma$ scatter above (the {\em Full} sample) and below ({\em HighL} sample) the intrinsic slope. 
 
\section{Cosmology and the circularity problem}

In previous approach \citep{DCC08,Dainotti2010,D11} some of us estimated the parameters $(a, b)$ and the intrinsic scatter $\sigma_{int}$ assuming the fiducial $\Lambda$CDM flat cosmological model with $\Omega_M = 0.291$ and $H_0 = 71  {\rm Km s}^{-1} {\rm Mpc}^{-1}$. We adopt the same cosmological model for simulating the data.
When these parameters, a and b, are fixed by a given cosmology we face the so called {\it circularity problem}.
 In order to determine the GRB luminosity $L_X^{*}$, we need to set a cosmological model, namely the determination of the calibration parameters $(a, b, \sigma_{int})$ can be different depending on which cosmology is adopted. Although several methods have been proposed to avoid this problem \citep{K08,L08,WZ08,CCD09,CI10,MEC11,MECP12}, it is highly desirable according to Petrosian et al. (2009) approach, to correctly take care of it in its full generality fitting together both the calibration parameters $(a, b, \sigma_{int})$ and the cosmological parameters each time for a given model.

Following \cite{DOC11}, we therefore constrain the ${\bf p}_{GRB} = (a, b, \sigma_{int})$ calibration quantities and the set of cosmological parameters ${\bf p}_c$ by considering the likelihood function\footnote{Note that we have here not expanded $\log{L_X^{*}}$ in terms of the measured quantities ${\bf p}_{obs}$ so that our likelihood function looks different from the one in \cite{DOC11}. The two expressions are actually consistent with each other should we use the relation among $L_X^{*}$ and ${\bf p}_{obs}$.}\,:

\begin{equation}
{\cal{L}}_{GRB} = \frac{1}{(2 \pi)^{{\cal{N}}_{GRB}/2}\Gamma_{GRB}^{1/2}({\bf p}_{GRB})} \
\exp{\left [ - \frac{\chi^2_{GRB}({\bf p}_{GRB}, {\bf p}_{c})}{2} \right ]}
\label{eq: deflikegrb}
\end{equation}
where

\begin{equation}
\chi^2_{GRB} = \sum_{i = 1}^{{\cal{N}}_{GRB}}{\frac{\log{L_X^{*}({\bf p}_c, {\bf p}_{obs}^{i})} - a \log{(T_a^{*})_i} - b}{a^2 (\sigma_T^i)^2 + (\sigma_L^i)^2 + \sigma_{int}^2}}
\label{eq: defchigrb}
\end{equation}

and

 \begin{equation}
\Gamma_{GRB}({\bf p}_{GRB}) = \prod_{i = 1}^{{\cal{N}}_{GRB}}{a^2 (\sigma_T^i)^2 + (\sigma_L^i)^2 + \sigma_{int}^2}
\label{eq: defgammagrb}
\end{equation}

with ${\rm p}_{obs}^i = (F_a^i, T_a^i, \beta^i)$ the set of simulated quantities needed to estimate $\log{L_X^{*}}$ for the $i$\,-\,th GRB given the cosmological parameters ${\bf p}_{c}$, $\sigma_T^i$ the error on $\log{T_a^{*}}$,$\sigma_L^i$ the one on $\log{L_X^{*}}$ obtained by propagating the measurement uncertainties on ${\bf p}_{obs}$, and the sum is over the ${\cal{N}}_{GRB}$ objects in the sample. 

Eqs.(\ref{eq: deflikegrb})\,-\,(\ref{eq: defgammagrb}) are the same as some of us have adopted in previous papers \citep{DCC08,Dainotti2010,D11} motivated by a Bayesian approach \citep{Dago05,Kelly} to the calibration problem with the only difference that the best fit zeropoint $b$ is not analytically expressed as a function of $(a, \sigma_{int})$, but it is free and it is added to the list of quantities to determine which thus sums up to ${\cal{N}}_c + 3$. In order to strengthen the constraints in such a large dimensional space, we add two further datasets, the Union2.1 SNIa sample containing 580 objects over the redshift range $0.015 \le z \le 1.414$ \citep{Union2.1} and the $H(z)$ over the redshift range $0.10 \le z \le 1.75$ \citep{S10I}. 
The combined likelihood for GRBs, SNeIa and $H(z)$ data simply reads: ${\cal{L}} = {\cal{L}}_{GRB} \times {\cal{L}}_{SNeIa} \times {\cal{L}}_H \times {\cal{L}}_0$
where the last term ${\cal{L}}_0$ with $(h_{obs}, \sigma_h) = (0.738, 0.024)$ has been introduced to take care of the recent measurements of the present day Hubble constant by the SHOES \citep{SHOES} collaboration.

For an assumed cosmological model characterized by a given ${\bf p}_c$ parameters set to be determined, the best fit calibration quantities $(a, b, \sigma_{int})$ will be the ones which maximize the full likelihood function ${\cal{L}}({\bf p}_{GRB}, {\bf p}_{c})$. In order to efficiently sample the ${\cal{N}}_c + 3$ dimensional parameter space, we use a Markov Chain Monte Carlo (MCMC) method running three parallel chains and using the \cite{GR92} test to check convergence. The histograms of the parameters from the merged chain after burn in cut and thinning are then used to infer median values and confidence ranges.

\section{Results}

In order to evaluate the likelihood function and hence constrain the cosmological parameters, we have first to choose a cosmological model. To this end, we will assume a two component universe filled by dust matter and dark energy (DE) with equation of state (EoS) given by the CPL \citep{CP01,L03} ansatz $w(z) = w_0 + w_a z/(1 + z)$. The dimensionless Hubble parameter $E(z) = H(z)/H_0$ then reads\,:

\begin{eqnarray}
E^2(z) & = & (1 - \Omega_M - \Omega_X) (1 + z)^2 + \Omega_M (1 + z)^3 \nonumber \\
 & + & \Omega_{X} (1 + z)^{3(1 + w_0 + w_a)} \exp{\left ( - \frac{3 w_a z}{1 + z} \right )}
\label{eq: ezfull}
\end{eqnarray}
with $(\Omega_M, \Omega_X)$ the present day matter and DE density parameters. Since we are mainly interested to show how the constrain on cosmological parameters change with a different correlation slope rather than constraining the cosmological parameters $(\Omega_M, \Omega_X, w_0, w_a, h)$ themselves, we will consider only two particular cases. For the first model (referred to as OLCDM), we assume DE is described by a cosmological constant term thus setting $(w_0, w_a) = (-1, 0)$, but leave open the possibility that the universe is not spatially flat. As a second case (dubbed in the following FCPL), we force the model to be flat (hence $\Omega_X = 1 - \Omega_M)$, but allow for a varying DE EoS and let the fit determining $(w_0, w_a)$.

\begin{table*}
\begin{center}
\resizebox{17cm}{!}{
\begin{tabular}{ccccccccccc}
\hline
~ & \multicolumn{5}{c}{GRB + $\omega_M$ + $H_0$} & \multicolumn{5}{c}{GRB + SNeIa + $H(z)$ + $H_0$} \\
\hline
$Id$ & $x_{bf}$ & $\langle x \rangle$ & $\tilde{x}$ & $68\% \ {\rm CL}$  & $95\% \ {\rm CL}$ & $x_{bf}$ & $\langle x \rangle$ & $\tilde{x}$ & $68\% \ {\rm CL}$  & $95\% \ {\rm CL}$ \\
\hline \hline
~ & ~ & ~ & ~ & ~ & ~ & ~ & ~ & ~ & ~  \\
$\Omega_M$ &0.241 &  0.264 & 0.255 & (0.203, 0.325) & 0.166, 0.395) &0.348 & 0.326 &0.327 & (0.284, 0.367) & (0.22, 0.40) \\
~ & ~ & ~ & ~ & ~ & ~ & ~ & ~ & ~ & ~  \\
$\Omega_{X}$ & 0.91& 0.932 & 1.065 & (0.529, 1.18) & (0.256, 1.2) & 0.912 & 0.831& 0.852 & (0.703, 0.926) & (0.64, 0.967) \\
~ & ~ & ~ & ~ & ~ & ~ & ~ & ~ & ~ & ~  \\
$h$ & 0.73 & 0.731 & 0.732 & (0.706, 0.756) & 0.683 0.778) & 0.736 &0.731 & 0.733 & (0.712, 0.746) & (0.69, 0.763) \\
~ & ~ & ~ & ~ & ~ & ~ & ~ & ~ & ~ & ~  \\
$a$ &-1.52&  -1.61& -1.613& (-1.76, -1.478) & (-1.88, -1.33) & -1.66 &-1.58 &-1.545 & (-1.74, -1.45) & (-1.92, -1.362) \\
~ & ~ & ~ & ~ & ~ & ~ & ~ & ~ & ~ & ~  \\
$b$ & 52.94& 53.27&53.24 & (52.76, 53.81) & (52.23, 54.24) & 53.43 & 53.13 & 53.05& (52.66, 53.78) & (52.29, 54.38) \\
~ & ~ & ~ & ~ & ~ & ~ & ~ & ~ & ~ & ~  \\
$\sigma_{int}$ &0.93 & 0.95 & 0.96& (0.88, 1.045) & (0.82, 1.15) & 0.90 &  0.97 &  0.97 & (0.89, 1.044) & (0.83, 1.12) \\
~ & ~ & ~ & ~ & ~ & ~ & ~ & ~ & ~ & ~  \\
\hline
~ & ~ & ~ & ~ & ~ & ~ & ~ & ~ & ~ & ~  \\
$\Omega_M$ & 0.241 &0.264 &0.255 & (0.203, 0.325) & (0.166, 0.395) &0.348 & 0.325 &0.327 & (0.284, 0.367) & (0.221, 0.397) \\
~ & ~ & ~ & ~ & ~ & ~ & ~ & ~ & ~ & ~  \\
$w_{0}$ & -0.574 & -0.626 & -0.575 & (-0.714, -0.519) & (-1.108, -0.503) & -1.16 & -1.12 & -1.13 & (-1.26, -0.944) & (-1.43, -0.79) \\
~ & ~ & ~ & ~ & ~ & ~ & ~ & ~ & ~ & ~  \\
$w_{a}$ &  1.40 & 0.314 & 0.345 & (-0.882, 1.38) & (-1.43, 1.48) & -0.336&-0.420 &-0.402 & (-0.632, -0.21) & (-0.822, -0.147) \\
~ & ~ & ~ & ~ & ~ & ~ & ~ & ~ & ~ & ~  \\
$h$ &0.75 &0.735 &  0.735 & (0.712, 0.757) &( 0.686, 0.787) &0.724 &0.728 &0.726 & (0.705, 0.751) & (0.689, 0.771) \\
~ & ~ & ~ & ~ & ~ & ~ & ~ & ~ & ~ & ~  \\
$a$ & -1.52 & -1.592 & -1.592 & (-1.76, -1.43) & (-1.88, -1.31) & -1.63 &-1.65 &-1.66 & (-1.77, -1.52) & (-1.98, -1.31) \\
~ & ~ & ~ & ~ & ~ & ~ & ~ & ~ & ~ & ~  \\
$b$ & 52.8 & 53.1 & 53.06 & (52.5, 53.70) & (51.6, 54.24) & 53.41 &53.4 &53.4 & (53.0, 53.9) & (52.2, 54.5) \\
~ & ~ & ~ & ~ & ~ & ~ & ~ & ~ & ~ & ~  \\
$\sigma_{int}$ &0.93 &0.964 & 0.96 & (0.881, 1.05) & (0.81, 1.14) & 0.95 & 0.99 & 0.995& (0.94, 1.06) & (0.87, 1.15) \\
~ & ~ & ~ & ~ & ~ & ~ & ~ & ~ & ~ & ~  \\

\hline
\end{tabular}}
\end{center}
\caption{Constraints on the cosmological and calibration parameters using the {\it Full} GRBs sample. Columns report best fit ($x_{bf}$), mean ($\langle x \rangle$) and median ($\tilde{x}$) values and the $68$ and $95\%$ confidence limits. Upper (lower) half of the table refers to the OLCDM (FCPL) model.}
\label{tab: fullgrbtab}
\end{table*}

Since both the correlation coefficient and a rough visual examination have shown that the GRBs in the {\it Full} and {\it HighL} samples follow different $L_X$\,-\,$T_a$ correlations, i.e. the values of $(a, b, \sigma_{int})$ for the two sets are not equal, we will discuss the results of the likelihood analysis for two distinct cases depending on which GRB sample is used to compute ${\cal{L}}_{GRB}$. Finally, we will also check whether the inclusion of non GRB data biases in some unpredictable way the calibration of the $L_X$\,-\,$T_a$ correlation by fitting the cosmological models to a modified likelihood function defined as: ${\cal{L}}_{only} = {\cal{L}}_{GRB} \times {\cal{L}}_0 \times {\cal{L}}_M$ where the term ${\cal{L}}_M$ with  $(\omega_{M}^{obs}, \sigma_M) = (0.1356, 0.034)$ represents a constrain on the physical matter density $\omega_M = \Omega_M h^2$ from the WMAP7 \citep{WMAP7} CMBR analysis. Note that we include this term in order to alleviate the degeneracy among the calibration and the cosmological parameters When fitting to GRBs\,+\,SNeIa\,+\,$H(z)$, we do not include the ${\cal{L}}_M$ since the SNeIa\,+\,$H(z)$ data play the role of constraining the background expansion thus leaving to GRBs data the task to calibrate the $L_X$\,-\,$T_a$ correlation.

\subsection{Full sample}

Table\,\ref{tab: fullgrbtab} summarizes the results of the likelihood analysis using the {\it Full} GRB sample both with and without the SNeIa and $H(z)$ data for the two cosmological models considered.

Let us focus first on the calibration parameters $(a, b, \sigma_{int})$. A straightforward comparison among the four different set of constraints makes it evident that the slope $a$, the zeropoint $b$ and the intrinsic scatter $\sigma_{int}$ are robustly determined notwithstanding which is the underlying cosmological model and the dataset used. In fact, although the best fit and median values are  different, the $68\%$ CL are always well overlapped. Notwithstanding the 5 sigma scatter in the correlation intrinsic slope the results here are consistent with our previous works \citep{DCC08,D11} where, however, we have only considered flat models. A qualitative consideration may help to explain why the calibration parameters do not dramatically depend on  the cosmological model. Indeed, since most of the GRBs are at high $z$, they are probing the matter dominated era. In such a regime, the dimensionless Hubble parameter in Eq.(\ref{eq: ezfull}) is mainly driven by the $\Omega_M (1 + z)^3$ term with the other contributions being partially smoothed out by the integration needed to get the luminosity distance. As a consequence, due to the large uncertainties on the simulated data, the details of the underlying cosmological are masked out in the calibration procedures. The signature of the cosmology will come out using a still larger sample with very small uncertainties on $(\log{T_a^{*}}, \log{L_X^{*}})$.

\begin{figure}
\includegraphics[width=8.5 cm]{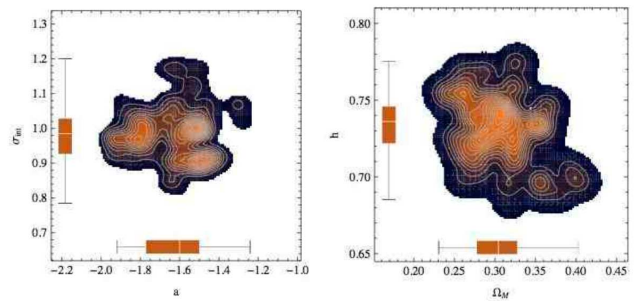}
\caption{ \textbf{Left panel} Regions of confidence for the marginalized likelihood function $ {\mathcal L}(a,\sigma)$, obtained marginalizing over $b$ and the cosmological  parameters using the total simulated sample. The bright brown regions indicate the $1\sigma$ (full zone) and $2\sigma$ (bright gray) regions of confidence respectively. On the axes are plotted the box-and-whisker diagrams relatively to the $a$ and
$\sigma_{int}$ parameters:  the bottom and top of the diagrams are  the 25th and 75th percentile (the lower and upper quartiles, respectively), and the band near the middle of the box is  the 50th percentile (the median). {\bf Right panel}  Regions of confidence for the marginalized likelihood function ${ \mathcal L}(\Omega_m, h)$.}
\label{conreg}
\end{figure}

We also underline that the departure from the \textit{real intrinsic }slope of the LT correlation affects indeed the cosmological paramaters.
First, we consider the OLCDM model which we have parameterized in terms of the present day values of the matter and DE density parameters and the Hubble constant. While $h$ is well in agreement with the estimates from both the local distance estimators \citep{SHOES} and the CMBR based data \citep{WMAP7}, the median values for both $(\Omega_M, \Omega_X)$ are  larger if compared to a fiducial $\Omega_M \sim 0.27$ obtained in previous works (see, e.g., \citealt{D07,Union2}).  Therefore, using a different intrinsic slope will bring a difference of the $13 \%$ on the best estimate of the $\Omega_M$ parameter, see Fig. \ref{conreg}. The constraints on the curvature parameter which turn out to be\,:

\begin{displaymath}
(\Omega_k)_{bf} =  -0.52\ \ , \ \ \langle \Omega_k \rangle =-0.24 \ \ , \ \ \tilde{\Omega}_k = -0.37 \ \ ,
\end{displaymath}

\begin{displaymath}
{\rm 68\% \ CL} = (-0.49, 0.15) \ \ , \ \ {\rm 95\% \ CL} = (-0.59, 0.45)
\end{displaymath}
for the fit to GRB\,+\,$\omega_M$\,+\,$H_0$ and

\begin{displaymath}
(\Omega_k)_{bf} = -0.305 \ \ , \ \ \langle \Omega_k \rangle =-0.177 \ \ , \ \ \tilde{\Omega}_k =-0.22 \ \ ,
\end{displaymath}

\begin{displaymath}
{\rm 68\% \ CL} = (-0.33, 0.0335) \ \ , \ \ {\rm 95\% \ CL} = (-0.39, 0.11)
\end{displaymath}
when fitting the GRB\,+\,SNeIa\,+\,$H(z)$\,+\,$H_0$ dataset. Here we even have median values point towards non flat models for both fits, a spatially flat universe is in agreement with, e.g., the WMAP7 only within the $95\%$ giving $\Omega_k = -0.080_{-0.093}^{+0.071}$.
This discrepancy can be outlined not only from the fact that in this case we are unable to strongly discriminate among flat and non flat models, but also from the fact that this is still not possible when SNeIa data are added to the fit. 

\begin{table*}
\begin{center}
\resizebox{17cm}{!}{
\begin{tabular}{ccccccccccc}
\hline
~ & \multicolumn{5}{c}{GRB + $\omega_M$ + $H_0$} & \multicolumn{5}{c}{GRB + SNeIa + $H(z)$ + $H_0$} \\
\hline
$Id$ & $x_{bf}$ & $\langle x \rangle$ & $\tilde{x}$ & $68\% \ {\rm CL}$  & $95\% \ {\rm CL}$ & $x_{bf}$ & $\langle x \rangle$ & $\tilde{x}$ & $68\% \ {\rm CL}$  & $95\% \ {\rm CL}$ \\
\hline \hline
~ & ~ & ~ & ~ & ~ & ~ & ~ & ~ & ~ & ~  \\
$\Omega_M$ & 0.332 & 0.311 & 0.309 & (0.248, 0.374) & (0.187, 0.438) &0.39& 0.3455 & 0.356 & (0.284, 0.401) & (0.218, 0.443) \\
~ & ~ & ~ & ~ & ~ & ~ & ~ & ~ & ~ & ~  \\
$\Omega_{X}$ & 0.912& 0.576 & 0.532 & (0.151, 1.042) & (0.106, 1.241) & 0.824 & 0.774 & 0.787 & (0.687, 0.860) & (0.584, 0.923) \\
~ & ~ & ~ & ~ & ~ & ~ & ~ & ~ & ~ & ~  \\
$h$ & 0.743 & 0.738 & 0.738 & (0.714, 0.762) & (0.692, 0.786) & 0.731 & 0.738 & 0.738 & (0.721, 0.759) & (0.700, 0.777) \\
~ & ~ & ~ & ~ & ~ & ~ & ~ & ~ & ~ & ~  \\
$a$ & -0.46 & -0.40 & -0.37 & (-0.61, -0.20) & (-0.82, -0.06) & -0.37 & -0.32 & -0.28 & (-0.51, -0.18) & (-0.64, -0.12) \\
~ & ~ & ~ & ~ & ~ & ~ & ~ & ~ & ~ & ~  \\
$b$ & 50.58 & 50.47 & 50.41 & (49.89, 51.09) & (49.48, 51.72) & 50.30 & 50.21 & 50.14 & (49.78, 50.68) & (49.62, 51.02) \\
~ & ~ & ~ & ~ & ~ & ~ & ~ & ~ & ~ & ~  \\
$\sigma_{int}$ & 0.33 & 0.38 & 0.37 & (0.31, 0.45) & (0.26, 0.55) & 0.35 & 0.38 & 0.37 & (0.29, 0.46) & (0.26, 0.59) \\
~ & ~ & ~ & ~ & ~ & ~ & ~ & ~ & ~ & ~  \\
\hline
~ & ~ & ~ & ~ & ~ & ~ & ~ & ~ & ~ & ~  \\
$\Omega_M$ & 0.235 & 0.235 & 0.235 & (0.171, 0.297) & (0.119, 0.359) & 0.294 & 0.310 & 0.313 & (0.256, 0.368) & (0.214, 0.408) \\
~ & ~ & ~ & ~ & ~ & ~ & ~ & ~ & ~ & ~  \\
$w_{0}$ & -0.71 & -0.86 & -0.78 & (-1.21, -0.56) & (-1.46, -0.47) & -0.97 & -1.02 & -1.03 & (-1.18, -0.78) & (-1.35, -0.75) \\
~ & ~ & ~ & ~ & ~ & ~ & ~ & ~ & ~ & ~  \\
$w_{a}$ & 0.73 & -0.01 & -0.12 & (-0.45, 0.56) & (-1.22, 1.38) & -0.62 & -0.78 & -0.79 & (-1.03, -0.54) & (-1.33, -0.45) \\
~ & ~ & ~ & ~ & ~ & ~ & ~ & ~ & ~ & ~  \\
$h$ & 0.732 & 0.738 & 0.738 & (0.714, 0.762) & (0.690, 0.786) & 0.737 & 0.728 & 0.730 & (0.710, 0.747) & (0.693, 0.763) \\
~ & ~ & ~ & ~ & ~ & ~ & ~ & ~ & ~ & ~  \\
$a$ & -0.53 & -0.37 & -0.37 & (-0.59, -0.15) & (-0.80, -0.03) & -0.67 & -0.67 & -0.64 & (-0.85, -0.54) & (-1.07, -0.39) \\
~ & ~ & ~ & ~ & ~ & ~ & ~ & ~ & ~ & ~  \\
$b$ & 50.67 & 50.37 & 50.37 & (49.73, 50.98) & (49.38, 51.59) & 51.30 & 51.25 & 51.17 & (50.86, 51.83) & (50.41, 52.38) \\
~ & ~ & ~ & ~ & ~ & ~ & ~ & ~ & ~ & ~  \\
$\sigma_{int}$ & 0.32 & 0.38 & 0.37 & (0.31, 0.45) & (0.26, 0.57) & 0.38 & 0.36 & 0.34 & (0.30, 0.43) & (0.28, 0.54) \\
~ & ~ & ~ & ~ & ~ & ~ & ~ & ~ & ~ & ~  \\
\hline
\end{tabular}}
\end{center}
\caption{Same as Table\,\ref{tab: fullgrbtab} but using the {\it HighL} GRBs sample.}
\label{tab: highlgrbtab}
\end{table*}

Forcing the model to be spatially flat but modeling the DE EoS with the CPL ansatz leads to the constraints for the FCPL model listed in the second half of Table\,\ref{tab: fullgrbtab}. There is a striking difference among the results for the GRB\,+\,$\omega_M$\,+\,$H_0$ and the GRB\,+\,SNeIa\,+\,$H(z)$\,+\,$H_0$ datasets. First, we note that the matter density parameter $\Omega_M$ is larger when GRBs and SNeIa are fitted together.

As already stressed, the constraints on $(w_0, w_a)$ are radically different for the GRB\,+\,$\omega_M$\,+\,$H_0$ and the GRB\,+\,SNeIa\,+\,$H(z)$\,+\,$H_0$ fits.

\subsection{HighL sample}

Let us now consider the fits using only the 18 GRBs in the {\it HighL} sample, namely a sample of GRBs with luminosity larger than the threshold value $(\log{L_X^{*}})_{th} = 48.7$. Actually, the value of $\log{L_X^{*}}$ depends on the cosmological model adopted to estimate the GRB distance so that, in principle, whether a GRB enters or not the {\it HighL} sample is a function of the unknown underlying cosmological parameters. As it is clear, maximizing ${\cal{L}}$ is the same as minimizing the sum of the single $\chi^2$ terms. Let us suppose that there are two sets of cosmological parameters ${\bf p}_c$ which give comparable values of $\chi^2_{SNeIa} + \chi^2_H$ so that the preferred one will be that with the lowest $\chi^2_{GRB}$. This happen for a subsample by imposing the cut $\log{L_{X,fid}^{*}} \ge 48.7$ with $\log{L_{X,fid}^{*}}$ the value estimated for a fiducial flat $\Lambda$CDM model with $(\Omega_M, h) = (0.266, 0.710)$. Although such a choice is somewhat arbitrary, we have checked a posteriori that the results (summarized in Table\,\ref{tab: highlgrbtab}) do not change if we change the fiducial model used for the selection of {\it HighL} GRBs.

As for the fits to the {\it Full} sample, we again find that the constraints on the calibration parameters only weakly depend on either the cosmological model or the dataset used (with or without SNeIa and $H(z)$ data). Indeed, although the best fit and median values change, the $68\%$ CL are in full agreement. We also note that adding the SNeIa and $H(z)$ data does not significantly improve the constraints on the calibration parameters. 

\begin{figure}
\includegraphics[width=7 cm]{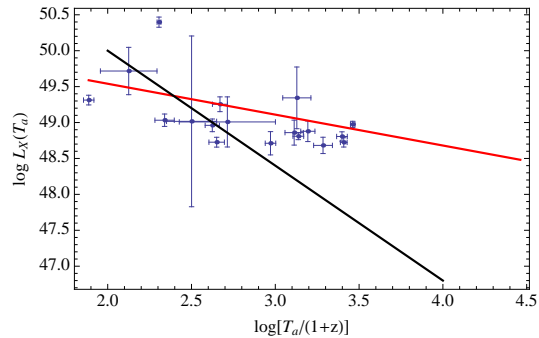}
\caption{Best fit curve for the  $ L_X^{*}$ -- $ T_a^{*}$ correlation relation (red line) superimposed on
the data in our  full bayesian approach, as it results for the FlatCPL model and GRB\,+\,SNeIa\,+\,$H(z)$\,+\,$H_0$ datasets. The best fit curve obtained for the full GRBdataset is superimposed (black line) just to highlight the difference in the slope.  }
\label{lxtaplothl}
\end{figure}

Comparing the calibrations parameters $(a, b, \sigma_{int})$ for the {\it Full} and {\it HighL} samples show us one of the most important outcome of the present analysis. The $L_X$\,-\,$T_a$ relation traced by the high luminosity GRBs is much more shallow (the median $|a|$ being smaller) and tight (with $\sigma_{int}$ decreasing from $\sim 1$ to $\sim 0.4$) than the corresponding one for the {\it Full} sample, see Fig. \ref{lxtaplothl}. A possible reduction of the intrinsic scatter for samples made out of large luminosity GRBs only was already pointed at in \cite{D11}. Here, some of us have shown that the selection on $\log{L_X^{*}}$ helps reducing the scatter without biasing in any other way the sample from the point of view of the other GRB properties (such as the time duration $T_a^{*}$ and the slope $\beta_a$ of the GRB energy spectrum). In that paper, we have, however, not explored further the dependence of the slope $a$ on the threshold luminosity, but only on the error energy parameter $\sigma(E)^2 = \sigma_{T_a^{*}}^2 + \sigma_{L_X^{*}}^2$. The more the sample is dominated by low $\sigma(E)$ GRBs, the shallower and tighter is the $L_X$\,-\,$T_a$ relation. Since low $\sigma(E)$ GRBs have typically large luminosity, the effect found in \cite{D11} goes in the same direction. This happens also with simulated data.

\begin{figure}
\includegraphics[width=8.5 cm]{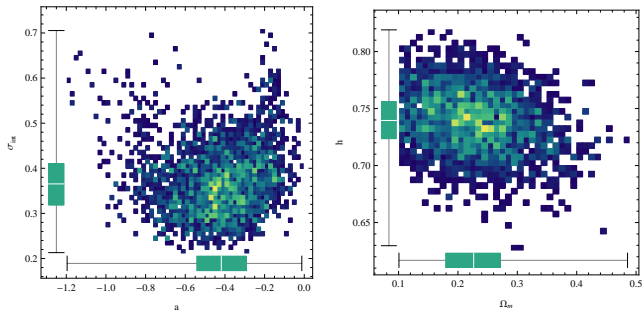}
\caption{ \textbf{Left panel} Regions of confidence for the marginalized likelihood function $ {\mathcal L}(a,\sigma)$, obtained marginalizing over $b$ and the cosmological  parameters for the high luminous sample. The bright brown regions indicate the $1\sigma$ (full zone) and $2\sigma$ (bright gray) regions of confidence respectively. On the axes are plotted the box-and-whisker diagrams relatively to the $a$ and
$\sigma_{int}$ parameters:  the bottom and top of the diagrams are  the 25th and 75th percentile (the lower and upper quartiles, respectively), and the band near the middle of the box is  the 50th percentile (the median). {\bf Right panel}  Regions of confidence for the marginalized likelihood function ${ \mathcal L}(\Omega_m, h)$.}
\label{conregh}
\end{figure}

We note that the {\it Full} sample results have closer value of the flat cosmological model predicted by the SNeIa sample, while the  {\it HighL} sample differs of $5\%$ in the value of $H_0$ computed in Peterson et al. 2010, while the scatter in $\Omega_M$ is underestimated by the $13 \%$, see Fig. \ref{conregh}. Therefore, we conclude that we need to follow one of these two approaches :
either we use an High luminous sample with the condition that  $\log{L_{X,fid}^{*}} \ge 48$, namely we have to chose the sample using as a fiducial cut exactly the luminosity for which the raw luminosity function coincides with the luminosity function corrected by the EP method \citep{Dainotti2013} or we should include in the evaluation of the cosmological parameters the luminosity and time evolution in the procedure described in sec. 3. 

\section{Conclusions}

The need to push the determination of the Hubble diagram deep into the matter dominated era has been the main driver for trying to make GRBs standardizeable candles so that they can probe cosmic expansion up to $z \sim 9.4$. The LT correlation we have discussed here stands out as one of the strategies implemented to achieve this goal emerging as the only one based on the X\,-\,ray afterglow quantities. Moving along the road explored in our previous works, we here show an appropriate analysis that one should do with GRB correlations considering their intrinsic slope and we show here how much changing the correlation slope we obtain departure from a standard cosmology.
The test with simulated data has also allowed us to jointly fit both the calibration and cosmological parameters, in order to fully take care of the circularity problem. To this end, we have added also SNeIa and $H(z)$ data to break the degeneracy among calibration and cosmology also checking whether the combination with non GRB data biases the estimate of the $(a, b, \sigma_{int})$ quantities.

The results of cosmology change using the high luminous or full sample, this means that the change of the correlation slope from the intrinisc one biases the cosmological results.
Here, we consider a simulated case, but the redshift evolution as it is explained in Dainotti et al. 2013 should be taken into account, in order to maximize the possibility to use correlations as cosmological tools. Moreover, in this approach the simulated data resemble a mixed sample of long GRBs and intermediate class (IC) \citep{nb2010}. Therefore, here we put another caveat on the use of whole sample of GRBs used for cosmological studies, namely being uncertain of the physical mechanism behind the LT correlation, one can not exclude the possibility that this unknown engine does not work in the same way for long and IC GRBs so that jointly fitting both classes biases the slope determination. Should this be the case, one would likely find a larger intrinsic scatter which is what we indeed get. Unfortunately, the number of IC GRBs in our sample is too small\footnote{{\it In the real sample we have only 11 IC GRBs as well as in the simulated sample}} to efficiently carry on the fitting analysis we have used here. Therefore, the validation of such hypothesis has to be postponed until a still larger dataset of IC GRBs is at our disposal. 

The existence of this sample suggests that the physical mechanism underlying the LT correlation is luminosity dependent, because above a certain luminosity value the raw luminosity function coincides with the luminosity function corrected by selection biases and redshift evolution.

In conclusion, we can claim that the present analysis opens a new perspective in the use of GRB correlations as cosmological tools, namely this research posed a strong caveat against the use of the observed GRB correlations not corrected by redshift evolution and selection biases. 

\section{Acknowledgments}
This work made use of data supplied by the UK Swift Science Data Centre at the University of Leicester. M.G.D. is grateful to Michal Ostrowski for the suggestion of considering the high luminous sample. M.G.D. is grateful for the support from Polish MNiSW through the Grant N N203 380336, the Fulbright Scholarship for its initial support and the Ludovisi- Blanceflor Foundation for the current support.


\begin{thebibliography}{99}

\bibitem[\protect\citeauthoryear{Amanullah et al. }{2010}]{Union2}
Amanullah, R., Lidman, C., Rubin, D., Aldering, G., Astier, P., et al. 2010, ApJ, 716, 712

\bibitem[\protect\citeauthoryear{Amati et al. }{2008}]{Amati08}
Amati, L., Guidorzi, C., Frontera, F., Della Valle, M., Finelli, F., Landi, R., Montanari, E. 2008, MNRAS, 391, 577

\bibitem[\protect\citeauthoryear{Bevington \& Robinson} {2003}]{Bevington}
Bevington, P. R., \& Robinson, D. K. 2003, Data Reduction and Error Analysis
for the Physical Sciences (3rd ed.; New York: McGraw-Hill)

\bibitem[\protect\citeauthoryear{Cardone et al. }{2009}]{CCD09}
Cardone, V.F., Capozziello, S., Dainotti, M.G. 2009, MNRAS, 400, 775

\bibitem[\protect\citeauthoryear{Cardone et al.} {2010}]{Cardone2010}
Cardone, V.F., Dainotti, M.G., et al. 2010, MNRAS tmp 1386C

\bibitem[\protect\citeauthoryear{Cardone et al. }{2011}]{Marcy}
Cardone, V.F., Perillo, M., Capozziello, S. 2011, MNRAS, 417, 1672

\bibitem[\protect\citeauthoryear{Capozziello \& Izzo }{2010}]{CI10}
Capozziello, S., Izzo, L. 2010, A\&A, 519, 73

\bibitem[\protect\citeauthoryear{Chevallier \& Polarski }{2001}]{CP01}
Chevallier, M., Polarski, D. 2001, Int. J. Mod. Phys. D, 10, 213

\bibitem[\protect\citeauthoryear{D' Agostini }{2005}]{Dago05}
D' Agostini, G. 2005, arXiv\,:\,physics/051182

\bibitem[\protect\citeauthoryear{Dainotti et al. }{2008}]{DCC08}
Dainotti, M.G., Cardone, V.F., Capozziello, S. 2008, MNRAS, 391, L79

\bibitem[\protect\citeauthoryear{Dainotti et al. }{2010}]{Dainotti2010}
Dainotti, M.G., Willingale, R., Cardone, V.F., Capozziello, S., Ostrowski, M. 2010a, ApJL,  722,  L215

\bibitem[\protect\citeauthoryear{Dainotti et al. }{2011}]{D11}
Dainotti, M.G., Cardone, V.F., Capozziello, S., Ostrowski, M., Willingale, R. 2011, ApJ, 730, 135

\bibitem[\protect\citeauthoryear{Dainotti et al.} {2011b}]{Dainotti2011b}
Dainotti, M.G., M. Ostrowski \& Willingale, R., 2011, MNRAS, 418, 2202D

\bibitem[\protect\citeauthoryear{Dainotti et al.} {2013}]{Dainotti2013}
Dainotti, M.G., Petrosian, V., Singal, J. \& M. Ostrowski, ApJ accepted

\bibitem[\protect\citeauthoryear{Davis et al. }{2007}]{D07}
Davis, T.M., M\"{o}rtsell, E., Sollerman, J., Becker, A.C., Blondin, S., et al. 2007, ApJ, 666, 716

\bibitem[\protect\citeauthoryear{Diaferio et al. }{2011}]{DOC11}
Diaferio, A., Ostorero, L., Cardone, V.F. 2011, JCAP, 10, 008

\bibitem[\protect\citeauthoryear{Demianski et al.}{ 2011}]{MEC11}
Demianski, M., Piedipalumbo, E., Rubano, C.,  2011, MNRAS, 411, 1213

\bibitem[\protect\citeauthoryear{Demianski et al. }{2012}]{MECP12}
Demianski, M., Piedipalumbo, E., Rubano, C. and Scudellaro, P. , 2012, MNRAS, 426: 1396Ð1415.

\bibitem[\protect\citeauthoryear{Efron \& Petrosian } {1992}]{Efron1992}
Efron, B. \& Petrosian, V., 1992, ApJ, 399, 345

\bibitem[\protect\citeauthoryear{Eisenstein et al. }{2005}]{Eis05}
Eisenstein, D.J., Zehavi, I., Hogg, D.W., Scoccimarro, R., Blanton, M.R. , et al. 2005, ApJ, 633, 560

\bibitem[\protect\citeauthoryear{Fenimore \& Ramirez\,-\,Ruiz }{2000}]{FRR00}
Fenimore, E.E., Ramirez\,-\,Ruiz, E. 2000, ApJ, 539, 712

\bibitem[\protect\citeauthoryear{Gelman \& Rubin }{1992}]{GR92}
Gelman, A., Rubin, D.B. 1992, Stat. Sci., 7, 457

\bibitem[\protect\citeauthoryear{Ghirlanda et al. }{2004}]{G04}
Ghirlanda, G., Ghisellini, G., Lazzati, D. 2004, ApJ, 616, 331

\bibitem[\protect\citeauthoryear{Ghisellini et al. }{2008}]{Ghisellini2008}
Ghisellini G., Nava L., Ghirlanda G., Firmani C., et al. 2008, A \& A, 496, 3, 2009.

\bibitem[\protect\citeauthoryear{Jimenez \& Loeb }{2002}]{JL02}
Jimenez, R., Loeb, A. 2002, ApJ, 573, 37

\bibitem[\protect\citeauthoryear{Kelly }{2007}]{Kelly}
Kelly, B.C. 2007, ApJ, 665, 1479

\bibitem[\protect\citeauthoryear{Kodama et al. }{2008}]{K08}
Kodama, Y., Yonetoku, D., Murakami, T., Tanabe, S., Tsutsui, R., Nakamura, T. 2008, MNRAS, 391, L1

\bibitem[\protect\citeauthoryear{Komatsu et al. }{2011}]{WMAP7}
Komatsu, E., Smith, K.M., Dunkley, J., Bennett, C.L., Gold, B. et al. 2011, ApJS, 192, 18

\bibitem[\protect\citeauthoryear{Kowalski et al. }{2008}]{Union}
Kowalski, M., Rubin, D., Aldering, G., Agostinho, R.J., Amadon, A. et al. 2008, ApJ, 686, 749

\bibitem[\protect\citeauthoryear{Lloyd \& Petrosian} {1999}]{Lloyd1999}
Lloyd, N., \& Petrosian, V. ApJ, 1999, 511, 550,

\bibitem[\protect\citeauthoryear{Liang \& Zhang }{2005}]{liza05}
Liang, E., Zhang, B. 2005, ApJ, 633, 611

\bibitem[\protect\citeauthoryear{Liang et al. }{2008}]{L08}
Liang, N., Xiao, W.K., Liu, Y., Zhang, S.N. 2008, ApJ, 385, 654

\bibitem[\protect\citeauthoryear{Linder }{2003}]{L03}
Linder, E.V. 2003, Phys. Rev. Lett., 90, 091301

\bibitem[\protect\citeauthoryear{Norris et al. }{2000}]{N00}
Norris, J.P., Marani, G.F., Bonnell, J.T. 2000, ApJ, 534, 248

\bibitem[\protect\citeauthoryear{Norris et al. } {2010}]{nb2010}
Norris, J.P. \& Bonnell, J.T. 2010, ApJ, 717, 411

\bibitem[\protect\citeauthoryear{Petersen et al. }{2010}]{Petersen2010}
Petersen, J. H., Holst K., \& Esben Budtz-Jørgensen, ApJ, 723, 966

\bibitem[\protect\citeauthoryear{Petrosian et al.} {2009}]{Petrosian2009}
Petrosian, V. Bouvier,A. \& Ryde, F. 2009, arXiv: 0909.5051P

\bibitem[\protect\citeauthoryear{Percival et al. }{2010}]{P10}
Percival, W.J., Reid, B.A., Eisenstein, D.J., Bahcall, N.A., Budavari, T. et al. 2010, MNRAS, 401, 2148

\bibitem[\protect\citeauthoryear{Piro }{2001}]{Piro01}
Piro, L. 2001, in {\it Gamma\,-\,ray Bursts in the Afterglow Era}, proceedings, Costa, E., Frontera, F. \& Hjorth, J. eds., Springer,Verlag, pp. 97

\bibitem[\protect\citeauthoryear{Riess et al. }{2009}]{SHOES}
Riess, A., Macri, L., Casertano, S., Sosey, M., Lampeitl, H., et al. 2009, ApJ, 699, 539

\bibitem[\protect\citeauthoryear{Schaefer }{2007}]{S07}
Schaefer, B.E. 2007, ApJ, 660, 16

\bibitem[\protect\citeauthoryear{Stern et al. }{2010a}]{S10I}
Stern, D., Jimenez, R., Verde, L., Kamionkowski, M., Stanford, S.A. 2010, JCAP, 02, 008

\bibitem[\protect\citeauthoryear{Stern et al. }{2010b}]{S10II}
Stern, D., Jimenez, R., Verde, L.,  Stanford, S.A., Kamionkowski, M. 2010, ApJS, 188, 280

\bibitem[\protect\citeauthoryear{Wei \& Zhang }{2009}]{WZ08}
Wei, H., Zhang, S.N. 2009, Eur. Phys. Journ. C, 63, 139

\bibitem[\protect\citeauthoryear{Suzuki et al.  }{2012}]{Union2.1}
Suzuki et al. (The Supernova Cosmology Project),  2012,  ApJ,  746, 85

\bibitem[\protect\citeauthoryear{Willingale et al. }{2007}]{W07}
Willingale, R.W., O' Brien, P.T., Osborne, J.P. et al. 2007, ApJ, 662, 1093

\bibitem[\protect\citeauthoryear{Xiao \& Schaefer }{2009}]{XS10}
Xiao, L., Schaefer, B.E., 2009, ApJ, 707, 387

\bibitem[\protect\citeauthoryear{Yamazaki }{2009}]{Yamazaki09}
Yamazaki, R. 2009, Apj, 690, L118

\end{thebibliography}
\end{document}